\documentclass[twocolumn,showpacs,amsmath,amssymb,showkeys,floatfix,prl]{revtex4}
\usepackage{amsfonts,amsmath}
\usepackage{graphicx}
\usepackage{dcolumn}
\usepackage{bm}

\begin{document}

\title{Calculating error bars for neutrino mixing parameters}

\author{H. R. Burroughs$^1$, B. K. Cogswell$^1$, J. Escamilla-Roa$^1$, D. C. Latimer$^2$, and D. J. Ernst$^1$}

\affiliation{$^1$Department of Physics and Astronomy, Vanderbilt University,
Nashville, Tennessee 37235}

\affiliation{$^2$Department of Physics, Reed College, Portland, Oregon 97202}

\begin{abstract}
One goal of contemporary particle physics is to determine the mixing angles and mass-squared differences that constitute
the phenomenological constants that describe neutrino oscillations. Of great interest are not only the
best fit values of these constants but also their errors. Some of the neutrino oscillation data is statistically poor
and cannot be treated by normal (Gaussian) statistics. To extract 
confidence intervals when the statistics are not normal, one should not utilize the value for $\Delta\chi^2$ versus confidence
level taken from normal statistics. Instead, we propose that one should use the normalized likelihood function as a probability distribution; the relationship between the correct $\Delta\chi^2$ and a given confidence level can be computed by integrating over the likelihood function. This allows for a definition of confidence level independent of the functional form of the $\chi^2$ function; it is particularly useful for cases in which the minimum of the $\chi^2$ function is near a boundary. 
We present two pedagogic examples and find that the proposed method yields confidence intervals that can
differ significantly from those obtained by using the value of $\Delta\chi^2$ from normal statistics. For example,
we find that for the first data release of the T2K experiment  the probability that $\theta_{13}$ is not zero, as defined by the maximum confidence level at which the value of zero is not allowed, is 92\%. Using the value of $\Delta\chi^2$ at zero
and assigning a confidence level from normal statistics, a common practice, gives the over estimation of 99.5\%.
\end{abstract}



\maketitle

Neutrino oscillations is the unique experimentally observed phenomenon that goes beyond the standard
model of the electroweak interaction.  Assuming the observations can be understood within the context of three neutrino flavors, a  coherent picture of the global data is sought in terms of two mass-squared differences, three mixing angles, and
one CP phase. 
In order to extract these parameters from the data, a model of each experiment is developed. The model results for the
experiment are then compared to the data through a choice of a particular statistic, often expressed as a  $\chi^2$
function. For a sufficiently large data set, 
normal (Gaussian) statistics can be
assumed, and the $\chi^2$ function is defined as:
\begin{eqnarray}
\chi^2(\{a_j\})&=:&\sum_{i}\,\frac{(n^\text{th}_i(\{a_j\},\{c_k\})-n^\text{exp}_i)^2}{\sigma_i^2}\nonumber \\
&&+\sum_k\,\frac{(c_k-c^\text{th}_k)^2}{\sigma_k^2}\,\,,
\label{normal}
\end{eqnarray}
where $\{a_j\}$ is a set of parameters, the mixing angles and mass-squared differences, to be determined; $\{c_k\}$ is a
set of systematic errors; $n^\text{exp}_i$ are
the experimental data points; $n^\text{th}_i(\{a_j\},\{c_k\})$  are the theoretical predictions of the data; $\sigma_i$ are the
statistical errors for the data  points;  $c^\text{th}_k$ are the best estimates of the systematic errors; and $\sigma_k$ the
errors for the systematics. The systematic error parameters are usually treated as nuisance parameters and $\chi^2(\{a_j\})$ is
minimized with respect to these parameters, often using the pull method \cite{Fogli:2002pt}, for each set of the parameters $\{a_j\}$. 
The best fit parameters are then the
values of the $a_j$ which minimize $\chi^2(\{a_j\})$.

Neutrino oscillations require that we must also deal with small statistical samples.  In particular, 
the recent T2K results \cite{Abe:2012gx} report a total of six observed neutrino events, binned by energy into sets containing zero, one, or two counts each. Despite this paucity, the data is a significant indicator that $\theta_{13}$ is non-zero. 
The Super-K atmospheric 
data afford another example.  Though it provides relatively stringent bounds upon the mixing angle $\theta_{23}$ and the ``atmospheric" mass-squared difference, the data also impact the determination of $\theta_{13}$.
The Super-K experiment provides an upper bound for the angle and shows a slight preference for negative values of
$\theta_{13}$  \cite{Roa:2009wp,Escamilla:2008vq}.   
The sensitivity of the data to $\theta_{13}$ can be traced to sub-GeV neutrinos with very long baselines \cite{Latimer:2004gz} and the MSW resonances that occur for normal hierarchy in the 3 to 7 GeV range \cite{Escamilla:2008vq}. The statistical
significance of the data in these two regions is low and the resulting $\chi^2$ is not well
represented by a quadratic so that the assumption of Gaussian statistics is tenuous. 

For small sample sizes, it is standard usage to employ a $\chi^2$ function defined in terms of Poisson statistics,
\begin{eqnarray}
\chi^2(\{a_j\})&=:&\sum_i\,2\,(n^\text{th}_i(\{a_j\},\{c_k\})+b_i-n^\text{exp}_i)\nonumber \\
&&+n^\text{exp}_i\log\left(\frac{n^\text{exp}_i}{n^\text{th}_i(\{a_j\},\{c_k\})+b_i}\right)\nonumber \\
&&+\sum_k\,\frac{(c_k-c^\text{th}_k)^2}{\sigma_k^2}\,\,,
\label{poisson}
\end{eqnarray}
where $b_i$ is a theoretical estimate of background events. 
The best fit parameters remain the values of the $a_j$ at the minimum value of $\chi^2(\{a_j\})$. In addition to being
valid for small sample sizes, this $\chi^2$ allows for the treatment of the situation where it is not possible to cleanly
separate the signal from the background. Background estimates are usually assessed through Monte Carlo simulations of the experimental detection and then inserted into Eq.~(\ref{poisson}). For large sample sizes, the Poisson
$\chi^2$ limits to the normal statistic $\chi^2$, thus allowing its use for data where some bins have good
statistics but some have poor statistics, as is the case for atmospheric data. 

Herein, we address the question as to how one should extract the errors on these parameters at a given
confidence level. 
A common practice is to use the value of $\Delta\chi^2 =: \chi^2 -\chi_{\text{min}}^2$ 
that corresponds to the desired confidence level 
as found from normal statistics, and then define the allowed
region for the parameter $a$ as lying within the interval $[a_o-\delta_1, a_o+\delta_2]$ where $\chi^2(a_o\pm \delta_{1,2}) = \chi^2_\text{min} + \Delta\chi^2$ with $a_o$ corresponding to the best fit.  For example, in a review on $\theta_{13}$ phenomenology \cite{Mezzetto:2010zi}, the authors quote the 90\% CL for $\sin^2 \theta_{13}$ computed by several groups.  As this mixing angle is small and the parametrization of the mixing angle is strictly positive, it is near zero, the boundary of the parameter space.  By observation, it is apparent that the $\chi^2$ for this parameter is manifestly not a quadratic and thus does not correspond to normal statistics.
The authors state that their quoted 90\% confidence levels on $\sin^2 \theta_{13}$ is 
found using the value $\Delta\chi^2=2.71$, but for the reasons cited above, caution must be employed in using this value.
Indeed, the authors of Ref.~\cite{Mezzetto:2010zi} admonish us that  ``the results on $\theta_{13}$ \dots  should be taken with some grain of
salt, and in particular the numbers given for various confidence levels \dots  have to be considered only as approximate, and
should always be understood in terms of the $\Delta\chi^2$ value." 

We propose a method for extracting allowed regions for a single parameter at a given confidence level that does not depend on the use of normal statistics. 
Instead, we take a Bayesian approach and interpret the normalized likelihood function with a flat prior as a probability distribution function. 
The likelihood function, $\mathcal L$ is defined in terms of the $\chi^2$ function by
\begin{equation}
 \chi^2(\{a_i\})  =: -2\,\log\,\mathcal L(\{a_i\}) \,.
\label{like}
\end{equation}
For a single parameter $a$, normal statistics give $\chi^2=(a-a_o)^2/\sigma^2$ and ${\mathcal
L}=\exp(-(a-a_o)^2/2\,\sigma^2)$, where $\sigma$ is the one standard deviation
error for $a$. 
For a compact parameter space, as with the mixing angles, one can assuredly normalize $\mathcal{L}$; for the mass-squared differences, the likelihood function falls off rapidly enough so that normalization is possible for these parameters as well.
We will hereafter work with a normalized likelihood function, $\overline{\mathcal L}$.

We begin with a brief summary of marginalization, as this leads directly to our proposal for determining error bars. 
Generally, the $\chi^2$ function and the maximum likelihood function are a function of $n$ parameters, $\{a_i\}$. Here these are the two mass-squared differences and the three mixing angles. Suppose we wish to extract information about one particular parameter, say $a_1$, in light of the knowledge of the remaining $n-1$ parameters. Marginalization tells us how to do so
\begin{equation}
\overline{\mathcal L}(a_1) = \int\, \mathrm{d}a_2\,\mathrm{d}a_3\,\dots \mathrm{d}a_n\, \overline{\mathcal L} (\{a_i\})\,\,.
\label{marg}
\end{equation}
This follows simply because the normalized likelihood function is a probability distribution function; hence, 
$\overline{\mathcal L}(a_1)$ is also a probability distribution function. 

\begin{figure}
\includegraphics*[width=3in]{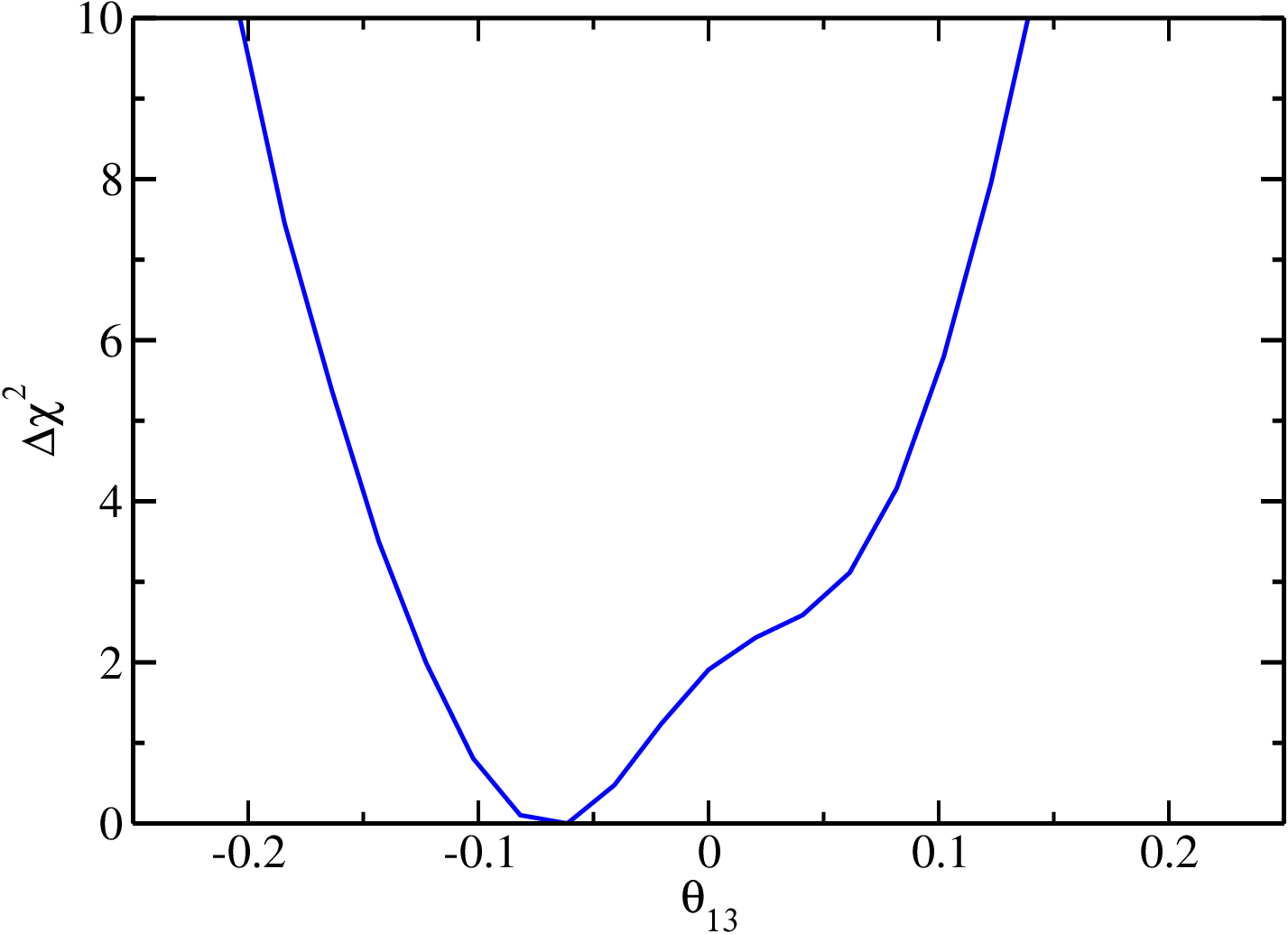}
\caption{$\Delta\chi^2$ versus the mixing angle $\theta_{13}$ as taken from the global analysis given in
Ref.~\protect\cite{Roa:2009wp}. }
\label{fig1}
\end{figure}

\begin{figure}
\includegraphics*[width=3in]{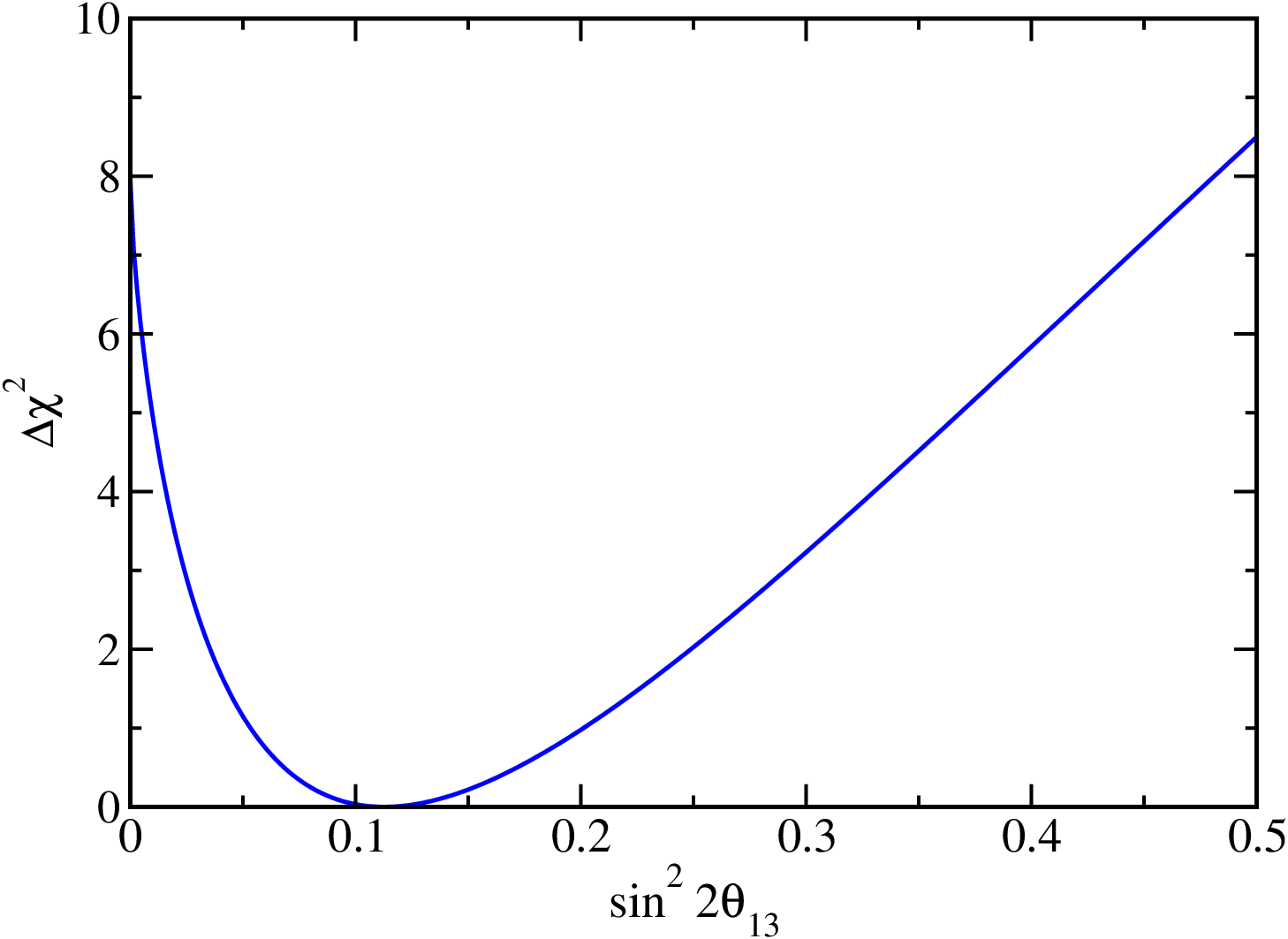}
\caption{$\Delta\chi^2$ versus $\sin^2\,2\theta_{13}$ for the T2K first data release \cite{Abe:2012gx} as taken from the
analysis in Ref.~\protect\cite{Cogswell:2012}. The curve depicted is calculated for positive $\theta_{13}$ and normal hierarchy. }
\label{fig2}
\end{figure}

 Dropping the subscript 1 for simplicity, we note that the probability ${\mathcal P}$ that the parameter $a$ lies between
$a_\text{min}$ and $a_\text{max}$  is 
\begin{equation}
{\mathcal P}(a_\text{min},a_\text{max}) =\int^{a_\text{max}}_{a_\text{min}}\,\overline{\mathcal L}(a)\,da\,\,.
\label{prob}
\end{equation}
We choose two pedagogic examples to demonstrate our results.

For Example 1, we consider the extraction of $\theta_{13}$ with $-\pi/2 \le \theta_{13}\le \pi/2 $ from the global analysis in Ref.~\cite{Roa:2009wp}.  [Note:  This analysis does not contain the recent data from Super-K III \cite{Wendell:2010md}, T2K \cite {Abe:2012gx},  MINOS neutrino disappearance \cite{Adamson:2011ig}, anti-neutrino dissappearance \cite{Adamson:2011ch} or neutrino appearance \cite{Adamson:2011qu}, Double Chooz \cite{Abe:2011fz}, or Daya Bay \cite{An:2012eh} experiments and is used here purely for illustrative purposes.]
In Fig.~\ref{fig1}, we plot $\Delta\chi^2$ versus $\theta_{13}$; note that $\Delta\chi^2$ is clearly not a
quadratic function. 
In Example 2, we consider the extraction  $\sin^22\,\theta_{13}$ from an analysis \cite{Cogswell:2012} of the
recent T2K data \cite{Abe:2012gx}. The T2K results are dependent on
the hierarchy and the sign of $\theta_{13}$; we show the results for normal hierarchy and positive
$\theta_{13}$. In Fig.~\ref{fig2}, we show $\Delta\chi^2$ versus $\sin^22\,\theta_{13}$.
Note that not only is $\Delta\chi^2$ not quadratic, but the minimum is near the lower bound of zero for
$\sin^2 2\,\theta_{13}$.

A simple application of Eq.~\ref{prob} would be to ask what is the probability calculated from  Fig.~\ref{fig1} that $\theta_{13}$ is less than zero. The result is 80\%. Similarly for Fig.~\ref{fig2} we can find that there is a 90\% probability that $\sin^22\,\theta_{13}\le 0.17$. 

To define a confidence level for the parameter $a$, we choose a value for $\Delta\chi^2$, find  the two points $a_o\pm \delta_{1,2}$  that correspond to the chosen
$\Delta\chi^2$, and integrate the likelihood function $\overline{\mathcal L}(a)$ from $a_o-\delta_1$ to $a_o+\delta_2$.
The integral yields the confidence level  associated with the particular value of $\Delta\chi^2$. If you desire a
particular confidence level, pick an initial guess for $\Delta\chi^2$, such as the value from normal
statistics, calculate the actual confidence level for this value and then repeat the process until you find the appropriate $\Delta\chi^2$ that produces the desired confidence level. The process is not computationally difficult nor
computationally intensive. Note that the concept of a standard deviation applies only to normal statistics, while
confidence level is universal.

\begin{figure}
\includegraphics*[width=3in]{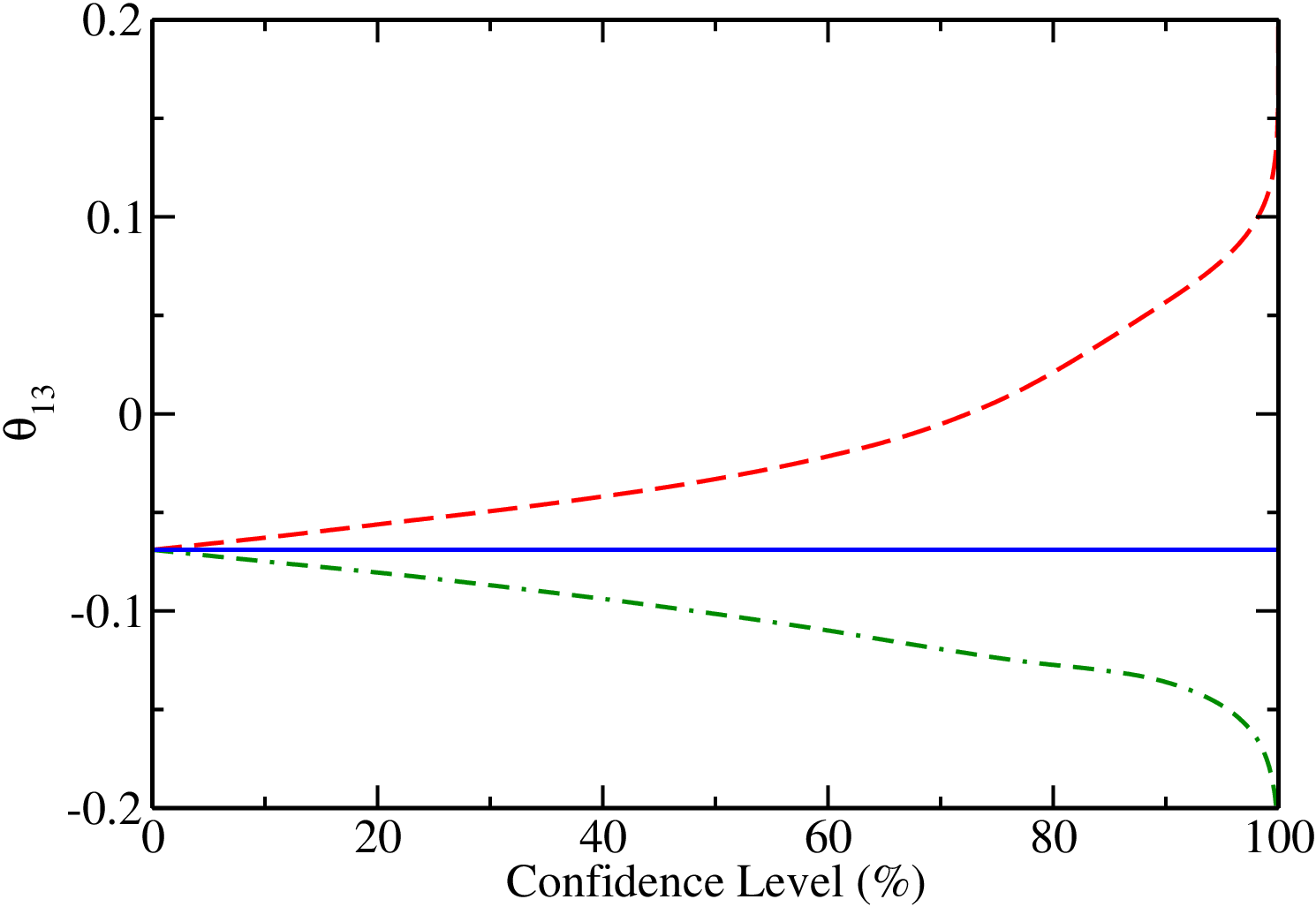}
\caption{[color online] The error bars as a function of confidence level for the $\Delta\chi^2$ from
Ref.~\protect\cite{Roa:2009wp} as depicted in
Fig.~\protect\ref{fig1}. The solid straight (blue) horizontal line is the minimum value of $\theta_{13}$, the dashed (red) line is
the upper end of the upper error bar while the dot-dash (green) curve is the lower end of the lower error bar. }
\label{fig3}
\end{figure}

\begin{figure}
\includegraphics*[width=3in]{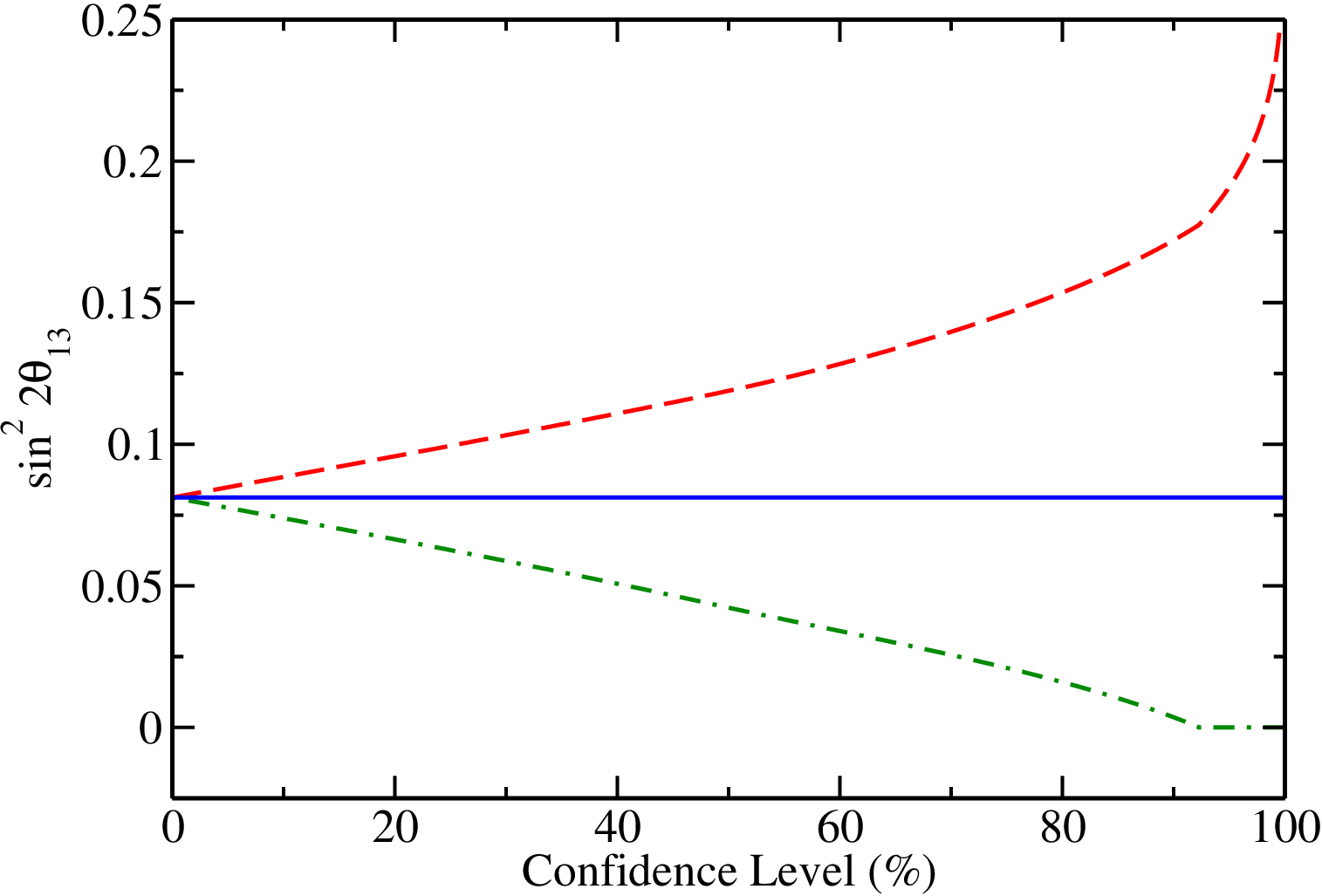}
\caption{[color online] The error bars as a function of confidence level for the $\Delta\chi^2$ for T2K
 \cite{Abe:2012gx} as depicted in
Fig.~\protect\ref{fig2}. The curves are the same as in Fig.~\protect\ref{fig3}. }
\label{fig4}
\end{figure}

For our two examples, we plot in Figs.~\ref{fig3} and \ref{fig4} the error bars on $\theta_{13}$ and $\sin^2 2\,\theta_{13}$, respectively,  as they vary with the confidence level.  Notice the errors are asymmetric in both cases.  In Fig.~\ref{fig4}, we see that the lower error bar for
$\sin^22\,\theta_{13}$  extends to zero and then remains there as the confidence level increases.
This demonstrates the point that, if the the best fit parameter is near a boundary of the parameter space, the confidence level will not be well approximated by the normal statistic, as $\Delta\chi^2$ is not  quadratic.

\begin{figure}
\includegraphics*[width=3in]{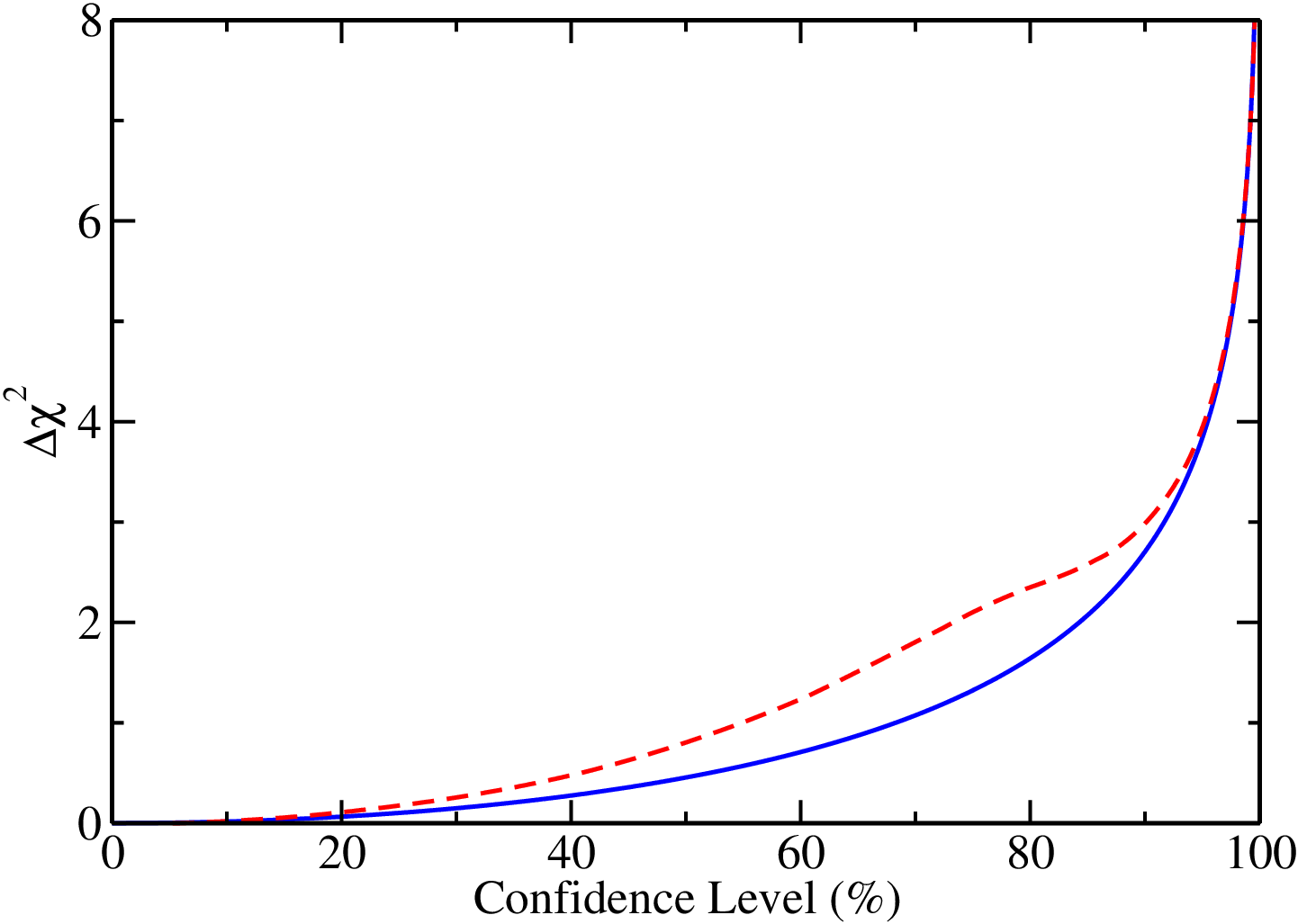}
\caption{[color online] The relationship of $\Delta\chi^2$ to the confidence level. The solid (blue) curve is for normal
statistics and the dashed (red) curve is calculated for the $\Delta\chi^2$ from the global analysis in
Ref.~\protect\cite{Roa:2009wp} 
as depicted in Fig.~\protect\ref{fig1}. }
\label{fig5}
\end{figure}

\begin{figure}
\includegraphics*[width=3in]{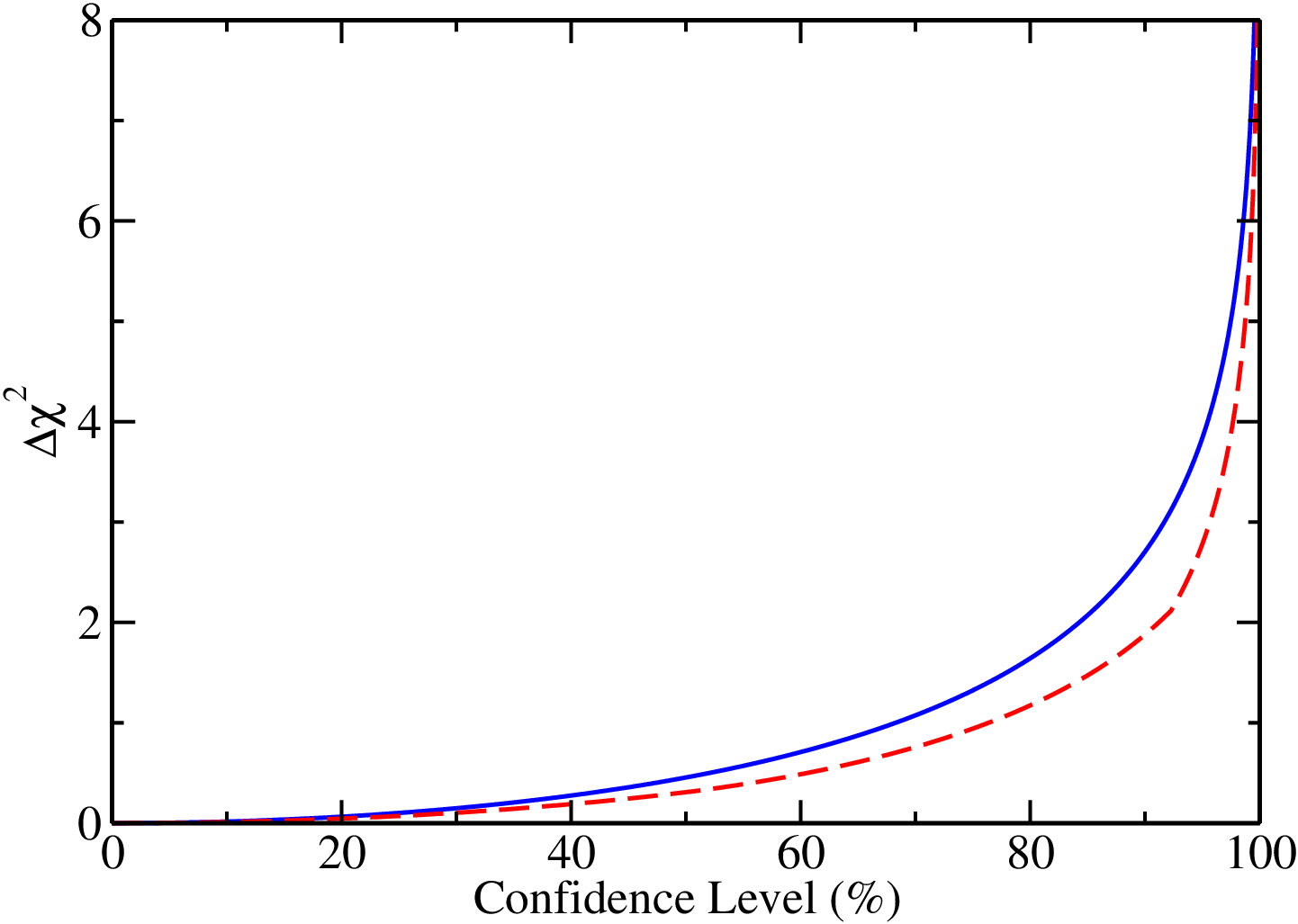}
\caption{[color online] The relationship of $\Delta\chi^2$ to the confidence level. The solid (blue) curve is for normal
statistics and the dashed (red) curve is calculated for the $\Delta\chi^2$ for the T2K experiment \cite{Abe:2012gx} in
Ref.~\cite{Cogswell:2012} as depicted in Fig.~\protect\ref{fig2} .}
\label{fig6}
\end{figure}

In Figs.~\ref{fig5} and \ref{fig6}, we examine the relationship between $\Delta\chi^2$ and the confidence level for our two examples, comparing our results with those from normal statistics.
In both figures, the [red] dashed curves utilize the normalized likelihood function, while the [blue] solid curves employ  normal statistics. In Table~\ref{tab1}, we present the same information for some commonly used confidence levels. We see that at low confidence levels there is a large difference between
either example and the normal statistics result. For example, from Table~\ref{tab1} we see that for Example 1 the 68\% confidence
level corresponds to a $\Delta\chi^2$ that is a factor of 1.7 larger than the normal statistics value of 1.00, and for Example
2 the $\Delta\chi^2$ is a factor of 0.7 lower than the normal statistics. For Example 1, we can understand why the correct
$\Delta\chi^2$ is larger than the normal statistics values up to the 99\% confidence level. This is because the
$\Delta\chi^2$  curve in Fig.~\ref{fig1} is more pointed than a quadratic, and it thus takes a higher value of
$\Delta\chi^2$ to get a
given percentage below that value. Also for Example 1, the correct and the normal statistics
value are nearly equal at a confidence level of 99\%, but this is accidental as the two confidence level curves intersect at a single point in this region. For Example 2, we see that the $\Delta\chi^2$ is always below the
normal statistics value. This feature will continue upward as the lower bound gets stuck at zero, and only  the upper
bound contributes above the chosen value of $\Delta\chi^2$, reducing that quantity by a factor of
approximately one half.

\begin{table}
\begin{tabular}{|c|c|c|c|}
\hline
~Confidence~ & \multicolumn{3}{c|}{$\Delta\chi^2$}\\ \cline{2-4}
Level (\% )     & Normal & ~~~Fig.~\protect\ref{fig1}~~~&~~~Fig.~\protect\ref{fig2}~~~ \\
           & ~Statstics~& $\theta_{13}$ & $\sin^22\,\theta_{13}$ \\
\hline
68.27  & 1.00 & 1.70 & 0.70 \\
\hline
90.00  & 2.71 & 3.00 & 1.88 \\
\hline
95.00  & 3.84 & 3.95 & 2.78 \\
\hline
95.42  & 4.00 & 4.09 & 2.93 \\
\hline
99.00  & 6.63 & 6.65 & 5.40 \\
\hline
99.73  & 9.00 & 8.90 & 7.55 \\
\hline
\end{tabular}
\caption{ The relationship of confidence level to $\Delta\chi^2$ for some commonly used confidence levels. Three
examples are given: 1.) normal statistics, 2.) the $\Delta\chi^2$ for $\theta_{13}$ taken from a global analysis \protect\cite{Roa:2009wp}
and shown in Fig.~\protect\ref{fig1}, and 3.) the $\Delta\chi^2$ for $\sin^2\,2\theta_{13}$ taken from 
Ref.~\protect\cite{Cogswell:2012} for the recent T2K data \cite{Abe:2012gx} and shown in Fig.~\protect\ref{fig2}.}
\label{tab1}
\end{table}

The question that remains to be answered is ``What is the probability that $\theta_{13}$ is or is not zero?" The correct answer
to this question is that the probability that $\theta_{13}=0$ is zero; the probability it is not zero is one.  Notice that
$\theta_{13}$ can be taken to lie between $-\pi/2$ and $+\pi/2$ and zero is a
single point out of the continuum. Thus the question is an ill-posed one. The more meaningful
question is ``What is the maximum confidence level at which zero is not an allowed value?" 
Consider Example 1, for normal statistics, we find $\Delta\chi^2(0)=2.0$ at $\theta_{13}=0$
so that we might claim that the mixing angle is nonzero at a confidence level of  84\%.
Using the likelihood function for Example 1, we find $\theta_{13}$ is non-zero at the 72\% confidence level, knowing that
we are here using the language somewhat loosely. 
For Example 2, the likelihood function excludes  
$\theta_{13}=0$ as an allowed value at the 92\% confidence level. The $\Delta\chi^2$ value at $\theta_{13}=0$ of 7.97 would give, using normal
statistics, 99.5\%. 
Why do we find this large over estimation? From Fig.~\ref{fig2} we see that below the minimum
$\Delta\chi^2$ rises quite rapidly while above the minimum $\Delta\chi^2$ rises slowly. This
combination will always yield an over estimation of the confidence level extracted from a single point on the lower,
rapidly rising curve. As the example case of T2K presented here is
typical, present claims which calculate the confidence level from normal statistics will overestimate the confidence that
zero is excluded from the allowed region for $\theta_{13}$.

In summary, we propose that confidence level and error bars be calculated based on the understanding that the
normalized likelihood function is a probability distribution function for whatever statistic is chosen to do the analysis. The
confidence level is then given by an integral over the normalized likelihood function, an implicit assumption in the marginalization procedure. 
We find that this alters the error bars we assign to
parameters, and that in the case of the minimum being near to an end point of the independent variable, such as in the
case of $\sin^22\,\theta_{13}$, the change that this procedure makes can be particularly significant. Further, we note that the
question of what is the probability that $\theta_{13}$ is not zero is more carefully worded as what is the maximum confidence
level at which the allowed region for $\theta_{13}$ does not include the value zero. Using this definition, published
confidence levels for non-zero $\theta_{13}$ based on the normal statistics relationship of $\Delta\chi^2$ to confidence
level are found to be over estimations.
\section{ACKNOWLEDGMENTS}

The work of B.~K.~C.~is supported, in part, by US Department of Education Grant P200A090275, the work of D.~J.~E.~is supported,
in part, by US Department of Energy Grant DE-FG02-96ER40975; the work
of J.~E-R. is supported, in part, by CONACyT, Mexico.

\bibliography{error.bib}

\end{document}